\documentclass[prl,aps,twocolumn,showpacs]{revtex4}
\begin{document}
\title{Effect of electron--phonon interaction on the
shift and attenuation of optical phonons}
\author{L. A. Falkovsky}
\affiliation{L.~D.~Landau Institute for Theoretical Physics, 2 Kosygin St., Moscow 117334, Russia}
\begin{abstract}
Using the Boltzmann equation for electrons in metals, we show that the optical
phonons  soften and have a dispersion
due to  screening in  agreement with the results
reported recently [M. Reizer, Phys. Rev. B {\bf 61}, 40 (2000)]. Additional phonon
damping and frequency shift arise when the electron--phonon interaction is properly 
included.
\end{abstract}
\maketitle
Despite attracting considerable interest for half a century since
the pioneering work by Fr\"ohlich, the problem of electron-phonon
interaction is still far from being solved. Migdal \cite{Mi}
developed a consistent many-body approach based on the Fr\"ohlich
Hamiltonian for interaction of electrons with acoustic (sound) phonons. As
Migdal showed ("the Migdal theorem"), the vertex corrections for
acoustic phonons are small by the {\it adiabatic} parameter
$\sqrt{m/M}$, where $m$ and $M$ are the electron and ion masses,
respectively. The theory described correctly the electronic
lifetime, renormalization of the Fermi velocity $v_F$ and acoustic
phonon attenuation but resulted in a strong renormalization of the
 sound velocity $\tilde{s}=s(1-2\lambda)^{1/2}$, where $\lambda$
is the dimensionless coupling constant. For sufficiently strong
electron-phonon coupling $\lambda \to 1/2$, the phonon frequency
approached to zero marking an instability point of the system.
Instead, one would
intuitively expect the phonon renormalization to be weak along
with the adiabatic parameter.

This discrepancy was resolved by Brovman and Kagan \cite{BK}
almost a decade later (see also \cite{G}).
They demonstrated the shortcomings of the Fr\"ohlich model that
gave an anomalously  large phonon renormalization.
Employing the Born--Oppenheimer
(adiabatic) approximation (see, e.g., \cite{BHK}),
they found that there are two terms in the second order
perturbation theory, which compensate each other
making a result
small by the adiabatic parameter. Namely, when calculating the
phonon self-energy function $\Pi (\omega,k)$ with help of the
diagram technique, one should eliminate an  adiabatic
contribution of the Fr\"ohlich model by subtracting $\Pi
(\omega,k)-\Pi(0,k)$.

The interaction of electrons with optical phonons was first
considered by Engelsberg and Schrieffer \cite{ES} within Migdal's
many-body approach for dispersionless phonons. They predicted a
splitting of the optical phonon  at finite wavenumbers $k$ into
two branches. Ipatova and Subashiev \cite{IS} calculated later on
the optical phonon attenuation in the collisionless limit and
pointed out that the Brovman-Kagan renormalization should be carried
out for optical phonons  in order to obtain correct phonon
renormalization.
 In the  paper
\cite{AS}, Alexandrov and Schrieffer corrected the calculational
error of Ref.\
\cite{ES} and argued that no splitting was found in fact. Instead,
they predicted an extremely strong dispersion of optical phonons,
 $\omega_k=\omega_0+\lambda v_F^2 k^2/3\omega_0$,
due to the coupling to electrons.
No such a dispersion has ever been observed
experimentally. The usual dispersion of optical phonons
in metals has the order of the sound velocity.
In a recent paper, Reizer \cite{R}  stressed the
importance of screening effects which should be taken into account.
The works \cite{AS}, \cite{R} are limited to the case of
collisionless both electron and phonon systems. Moreover, only
the phonon renormalization was considered with
no results available for the attenuation of optical
phonons.

A different from many-body technique  semiclassical approach
 based on the Boltzmann equation and the equations of the
elasticity theory  was developed in the papers by Akhiezer,
Silin, Gurevich, Kontorovich, and many others (we refer the reader
to the review \cite{Kon}). This approach
was compared with various experiments, such as
attenuation of sound waves, effects of
strong magnetic fields,   crystal anisotropy, and sample surfaces
on the sound attenuation, and so
on. It can be applied to the problem of the
electron--optical-phonon interaction  \cite{MF} as well. In the
present paper we develop a theory for both the attenuation and
frequency shift of optical phonons with  account
for effects of the
Coulomb screening as well as collisions in the electron and phonon
systems.

It is instructive to recapitulate the results of the dynamical theory
of elasticity  for
the renormalization of the sound velocity $\delta s=\tilde{s}-s$ and 
acoustic attenuation $\Gamma$
in metals. For a phonon with a wavenumber $k$ and frequency $\omega_k=sk$,
they are given by \cite{Kon}
\begin{eqnarray} \label{rsd}
\frac{\delta s}{s}-i\frac{\Gamma}{\omega_k}= \lambda
\left\{\begin{array}{ll}
\displaystyle
{s^2\over v_F^2} -i{\pi s\over 2v_F}& \mbox{ for $kv_F>|\omega_k+i\gamma|$},\\
\displaystyle
 \frac{\omega_k}{\omega_k+i\gamma}&
\mbox{ for $kv_F<|\omega_k+i\gamma|, $}
\end{array}\right.
\end{eqnarray}
where $\gamma$
is the electronic scattering rate,
and the dimensionless coupling constant
$\lambda $
is proportional to
the electronic
density of states $\nu_0$ at the Fermi surface (for the isotropic
case $\nu_0 = m^* p_F/\pi^2$, $m^*$ is the effective electron mass)
and  to the squared
deformation potential $\zeta_{ik}$. The deformation potential
describes the change in the spectrum of electrons subject to
lattice deformation
$ \varepsilon ({\bf p},{\bf r},t)=
\varepsilon _{0}({\bf p})
+\zeta_{ik}({\bf p})u_{ik}({\bf r},t)$,
where $u_{ik}$ is the strain tensor.
Equations (\ref{rsd}) give the correct answers in  various known
regimes: for the sound attenuation in the hydrodynamic limit
($\omega_k\ll \gamma$ and
$k\to 0$), for the zero-sound ($\omega_k\gg \gamma$ and $ k\to 0$),
for the Landau
damping in the ballistic limit ($kv_F\gg|\omega_k+i\gamma|$).
In the latter case,
both the sound velocity shift and attenuation
are small by the  adiabatic parameter
$s/v_F$ contrary to the results of the Fr\"ohlich
model. Note also that Eqs. (\ref{rsd}) show hardening of the phonon
frequency due to the electron--phonon interaction in contradiction with
Migdal's result.

For the case of optical phonons, two known types of
 the electron--phonon interaction  are
the deformation potential and  the interaction
with the electrical polarization induced by  optical vibrations.
We consider here for simplicity
a  cubic crystal with two different atoms in a unit cell. Then,
there are three optical modes, and
the interaction with the induced polarization has the Fr\"ohlich form:
\begin {equation} \label{spo}
\varepsilon ({\bf p},{\bf r},t)= \varepsilon _{0}({\bf p})
+\zeta({\bf p}) {\bf \nabla}\cdot {\bf u}({\bf r},t)
\end {equation}
where  the scalar function $\zeta({\bf p})$   of the electron momentum
is the coupling with the optical displacements ${\bf u}$.
In order to compare our results with previous ones, here we consider 
the interaction in the same form (\ref{spo}) as in Refs. \cite{AS}, \cite{R}.
One can see that  principal characteristic features of the phenomenon
are retained for the deformation interaction,
 $\delta\varepsilon=\zeta_i({\bf p})u_i({\bf r}, t)$,
where the coupling is a vector function.
The distinction is that only   the longitudinal mode interacts with
electrons in the case  of induced polarization (\ref{spo}), and the
interaction approaches zero in the long-wave limit.
Therefore, we  concentrate
on the propagation of the longitudinal mode along the symmetry axis when this
mode is not mixed with  transverse ones.
Note also that the electric field plays an important role especially
when the different atoms are in the unit cell so that the
dipole moment is excited under the atom vibrations.
At last, the optical phonons have always
the so-called natural width $\Gamma^{\text{nat}}\sim \omega_0\sqrt{m/M}$.
The natural width results from  decay
processes into two (or more) acoustic or optical phonons, which are possible
even at zero temperature.

The main point of the theory is the equation of  motion
in the long-wave approximation
($k\ll 1/a$) for the Fourier components of the
optical-phonon displacement $u_j$:
\begin{eqnarray}
\label{oeq}
(\omega_k^2-i\omega\Gamma^{\text{nat}}-\omega ^{2})u_{j}({\bf k},\omega)
=\frac{Z}{M'}E_j
\\ \nonumber+\frac{ik_j}{M'N}\int {2d^{3}p\over (2\pi )^{3}}
\zeta({\bf p})\delta f_{p}({\bf k},\omega),
\end{eqnarray}
where
$E_j$ is the electric field associated with vibrations,
$N$ is the number of unit cells in 1 cm$^3$,
$M'$ is the reduced mass of two atoms in the unit cell,
and  $Z$ is the  effective ionic charge.
The nonperturbed phonon frequency $\omega_k$ should be considered
in  the absence of
the electric field and without any nonadiabatic corrections.
In the long-wave limit, we can roughly  describe it as
$\omega_k^2=\omega^2_0\pm s^2 k^2$
with the magnitude of $s$ being of the order of the typical sound velocity
in metals.
The last term in Eq. (\ref{oeq}) presents the driving force from
the nonadiabatic electron system due to the deviation
$\delta f_{p}({\bf r}, t)$ from the local-equilibrium distribution
function $f_0[\varepsilon({\bf p},{\bf r},t)-\mu]$.

Then, we have  the Boltzmann equation
\begin{eqnarray}\label{be1}
-i(\omega-{\bf k}\cdot{\bf v})\delta f_{p}({\bf k},\omega)+
\gamma\left[\delta f_{p}({\bf k},\omega)-
\langle\delta f_{p}({\bf k},\omega)\rangle\right]\\ \nonumber
=-[\omega \zeta({\bf p}){\bf k}\cdot{\bf u}({\bf k},\omega)
+e{\bf v}\cdot{\bf E}]\frac{df_0}{d\varepsilon},
\end{eqnarray}
in the approximation of relaxation rate $\gamma$,
which holds at low temperatures
when the electron--impurity interaction dominates as well as at temperatures
higher then the Debye temperature
when the phonon--phonon collisions can be considered as elastic.
The term in the angle brackets in Eq. (\ref{be1}), which
denote the average over the Fermi surface,
\begin{displaymath}
\langle...\rangle = \frac{1}{\nu_0}\int(...){2dS_{F}\over v(2\pi)^3},
\end{displaymath}
arises from the out-term in the collision integral.
Notice, that
the condition
$\langle \zeta({\bf p})\rangle=0$
should be fulfilled, because the number of electrons
in the local-equilibrium state  $f_0[\varepsilon({\bf p},{\bf r},t)-\mu]$
is conserved.
With the help of the Maxwell equations, the electric field is
expressed in the terms of polarization $\bf P$ as follows:
\begin{equation} \label{me}
{\bf E }= -4\pi {\bf k} ({\bf k}\cdot{\bf P})/k^2,
\end{equation}
provided that phonons are excited in the optical region 
$k\gg\omega/c$ where the wavevector is determined by the incident light
$k\sim\omega^{(i)}/c$ and the  frequency $\omega$ is of the order of the
optical phonon frequency $\omega_0$.
 It is seen that the  electric field is longitudinal and
only the  longitudinal component of polarization $P_z$
(${\bf k}$ is taken along the $z$-axis) plays a role being
related to the phonon displacement and the electric field
by the equation
\begin{equation} \label{pol}
 P_z=NZ u_z+\alpha E+\frac{ie}{k}
\int\frac{2d^3p}{(2\pi)^3}
\delta f_p({\bf k},\omega),
\end{equation}
where the first term is caused by ionic motion,
$\alpha$ is the polarizability of  filled bands,
and the last term is the carrier contribution 
defined by the variation of the electron density,
$\rho^{(e)}=-i{\bf k}\cdot {\bf P}^{(e)}$.

Equations (\ref{oeq})--(\ref{pol}) give
the complete system of our problem.
The Boltzmann equation (\ref{be1}) has the solution in the form
$$\delta f_p({\bf k}, \omega)=-
\chi_p({\bf k}, \omega)
\frac{df_0}{d\varepsilon},$$
where
\begin{eqnarray} \label{sbe1}
\chi_p({\bf k}, \omega)
=
i[e{\bf v}\cdot{\bf E}+\omega\zeta({\bf p}) {\bf k}\cdot{\bf u}+\gamma
\langle\chi_p({\bf k}, \omega)\rangle]/\Delta,\\ \nonumber
\langle\chi_p({\bf k}, \omega)\rangle
=
i\langle [ e{\bf v}\cdot{\bf E}
+\omega \zeta({\bf p}){\bf k}\cdot{\bf u}]/\Delta
\rangle/(1-i\langle\gamma/\Delta\rangle),
\end{eqnarray}
and we set $\Delta=\omega-{\bf k}\cdot{\bf v}+i\gamma$.

Using this solution, we obtain the polarization (\ref{pol}) and rewrite
the electric field  (\ref{me}) in terms of the longitudinal displacement
$u_z$:
\begin{equation} \label{elf}
\varepsilon_e(k,\omega) E=-4\pi \beta_{fd}u_z,
\end{equation}
where we introduce the field--displacement response function
\begin{equation} \label{be}
\beta_{fd}=NZ -e\omega \nu_0\frac
{\langle\zeta({\bf p})/\Delta\rangle}
{1-i\langle \gamma/\Delta\rangle}.
\end{equation}
The electron contribution into the dielectric function has the known form:
\begin{equation}\label{df}
\varepsilon_e(k,\omega)-\varepsilon_{\infty}=
-\frac{4\pi e^2\nu_0
\langle v_z/
\Delta\rangle}{k(1-i\langle \gamma/\Delta\rangle)},
\end{equation}
where the high-frequency permittivity $\varepsilon_{\infty}= 1+4\pi\alpha$.

Now, we consider the equation of motion (\ref{oeq}) using
the solution of the Boltzmann equation (\ref{sbe1}).
The term proportional to ${ u_z}$ of the driving force can be included
in the phonon frequency:
\begin{equation}\label{sh}
\tilde{\omega}^2=\omega_k^2 -i\omega\Gamma^{\text{nat}}
+\frac{\nu_0\omega k^2}{M'N}\left(
\big\langle \frac{\zeta^2({\bf p})}{\Delta}\big\rangle
+\frac{i\gamma
\langle \zeta({\bf p})/\Delta\rangle^2 }
{1-i\langle\gamma/\Delta\rangle}
\right),
\end{equation}
so that  Eq.  (\ref{oeq}) reads
\begin{equation} \label{oeq1}
(\tilde{\omega}^2-\omega^2)u_z=
\tilde{Z}E/M',
\end{equation}
where the renormalized ionic charge
$$\tilde{Z}=Z-\frac{e\nu_0 k}{N}
\left(\big\langle\frac{\zeta({\bf p}) v_z}{\Delta}\big\rangle
+i\gamma\frac{\langle v_z/\Delta\rangle
\langle\zeta({\bf p})/\Delta\rangle}
{1-i\langle\gamma/\Delta\rangle}\right).$$
Using the condition
$\langle \zeta({\bf p})\rangle=0$ we obtain
$\tilde{Z}=\beta_{fd}/N.$

Then, we can express the displacement $u_z$ from Eq. (\ref{oeq1})
in terms of $E$ and, substituting into Eq. (\ref{elf}),
obtain the dielectric function of the electron--ion system:
\begin{equation}\label{dfu}
\varepsilon(k,\omega)=\varepsilon^{(e)}(k,\omega)+4\pi N\tilde{Z}^2/M'
(\tilde{\omega}^2-\omega^2).
\end{equation}

The frequency of the longitudinal mode, $\omega=\omega_{\text{LO}}$, is
defined by the condition
$\varepsilon(k,\omega)=0$, i.e.,
\begin{equation}\label{dfu1}
\omega^2=
\tilde{\omega}^2+4\pi N\tilde{Z}^2/M'\varepsilon_{e}(k,\omega).
\end{equation}

In  the absence of free electrons, the density of states 
$\nu_0=0$ and  Eq. (\ref{dfu1}) gives
for the LO mode
$\omega^2_{\text{LO}}=\omega_k^2+\omega_{pi}^2 - i\omega_{\text{L0}}\Gamma^{\text{nat}}$,
where $\omega_{pi}^2=4\pi NZ^2/\varepsilon_{\infty}$ is the squared
ion-plasma frequency of the order of $\omega_0^2$.
For the TO mode, when the electric field $E=0$, we obtain $\omega^2_{\text{TO}}=
\omega_k^2 - i\omega_k\Gamma^{\text{nat}}$.

Free electrons in metals make the large contribution into the
dielectric function [see, Eq. (\ref{df})].
Expanding in powers of $k$ we have in the zero-order 
\begin{equation} \label{Dr}
\varepsilon^{(e)}(k,\omega)
-\varepsilon_{\infty}=
-\varepsilon_{\infty}
\omega_{pe}^2/\omega(\omega+i\gamma),
\end{equation}
which corresponds simply to the Drude conductivity with the electron-plasma frequency
$$\omega_{pe}^2=\frac{e^2}{3\pi^2
\varepsilon_{\infty}}\int vdS_F.$$
For large
$kv_F>|\omega+i\gamma|$, the electron contribution (\ref{df}) 
describes the Debye screening:
$$\varepsilon^{(e)}(k,\omega)-\varepsilon_{\infty}=
\varepsilon_{\infty}
{k_0^2\over k^2}(1+i\pi\omega/2 kv_F),$$
where a term of the order of $\omega/kv$ is kept and the Debye 
parameter 
$k_0^2=4\pi e^2\nu_0/\varepsilon_{\infty}$.

Therefore, we can solve Eq. (\ref{dfu1}) for $\omega_{LO}\ll\omega_{pe}$,
using the iteration procedure. To a first approximation, we have
\begin{equation}\label{dfu2}
\omega^2=\omega_k^2 -i\omega_k\Gamma^{\text{nat}}+
\frac{k^2\omega_k\nu_0}{M'N}
\big\langle\frac{\zeta^2({\bf p})}{\Delta}\big\rangle
+{\omega_{pi}^2\varepsilon_{\infty}\over\varepsilon^{(e)}(k,\omega_k)}.
\end{equation}

In the case of  small $kv_F< |\omega_k+i\gamma|$,
expanding  in $k$-powers, we obtain
the solution
\begin{equation}\label{sol}
\omega^2_{\text{LO}}=\omega_k^2 -i\omega_k\Gamma^{\text{nat}}-
\omega_k(\omega_k+i\gamma)\frac{\omega_{pi}^2}{\omega^2_{pe}}
+\frac
{\lambda\omega_ks^2}{\omega_k+i\gamma}k^2,
\end{equation}
where the dimensionless coupling constant $\lambda 
=\langle\zeta^2 ({\bf p})\rangle\nu_0 / \rho s^2$  contains the factor
$\ ap_Fm^*/m$ and the metal density $\rho$.

In the case of large $k$, expanding in  $|\omega_k+i\gamma|/ kv_F$,
 we obtain
\begin{equation}\label{sol1}
\omega^2_{LO}=\omega_k^2 -i\omega_{k}\Gamma^{\text{nat}}-i
{\pi  \omega_k s^2k\over 2v_F}
\left(\lambda +\frac{\omega_{pi}^2}{s^2k_0^2}\right)
+\frac{\omega_{pi}^2}{k_0^2}k^2,
\end{equation}
where the  coupling constant $\lambda$ is defined, when the asymptotic
value of the integral is calculated:
$\nu_0\langle\zeta^2({\bf p})/\Delta\rangle/M'N=-i\pi s^2\lambda/2kv_F$.
Note, that the value of $\lambda$ vanishes in the isotropic case 
due to the condition $\langle\zeta({\bf p})\rangle=0$.

Hence, the squared frequency of the longitudinal optical mode is essentially
less (by the factor $\omega_{pi}^2$) than for insulators,  due to
screening of the electric field  by free electrons. The additional
phonon softening, width, and dispersion in
Eq. (\ref{sol}) involve the adiabatic
parameter $(\omega_0/\omega_{pe})^2\sim m/M$, and
they are small  compared
with $\Gamma^{\text{nat}}\sim \omega_0\sqrt{m/M}$.
In the region, where Eq. (\ref{sol1}) is valid, we see the additional
$k$-dependent width (terms in the parentheses),
which is comparable with $\Gamma^{\text{nat}}$. Here, the $\lambda$-term
conditioned by the electron--phonon interaction is similar to the
damping of the acoustic phonons [see the first formula in Eqs. (\ref{rsd})].
Now we omit  the shift containing the  small factor $(s/v_F)^2$.
The second term in the parentheses as well as
 the last term in Eq. (\ref{sol1}), giving
the $k-$dispersion, are induced  by  screening and $\lambda-$independent.
Since $\omega_{pi}^2/k_0^2\simeq s^2$,
this dispersion has the   typical value for the phonon branches.

Let us rewrite in our notations the respective results
of Ref. \cite{R}, Eqs. (9) and (7), retaining only  main terms:
\begin{equation}         \label{R1}
\omega^2_{\text{LO}}=\omega^2_k-\frac{\omega_{pi}^2(\omega_{pi}^2+\omega_k^2)}
{\omega_{pe}^2(1-2\lambda\ln2)}
,\qquad kv_F<\omega_k,
\end{equation}
\begin{equation} \label{R2}
\omega^2_{\text{LO}}=\omega^2_k+\frac{\omega_{pi}^2}{3\omega_{pe}^2}(kv_F)^2
,\qquad kv_F>\omega_k.
\end{equation}
Comparing with Eqs. (\ref{sol})-(\ref{sol1}) we see that
the $k$-dispersion  coincides practically.
Next, we agree  that the contribution $\omega^2_{pi}$
vanishes from the frequency of the LO mode.
The  softening and damping due to both
the electron--phonon  scattering ($\gamma$) and the
phonon decay-processes ($\Gamma^{\text{nat}}$)   were ignored
in Ref. \cite{R}.
Concerning the electron--phonon interaction $\lambda$,
the reason of disagreement was discussed in the beginning of the paper: this is
shortcomings of diagram technique based on the Fr\"ohlich model.
But the most essential difference is the $k$-dependent width in Eq.
(\ref{sol1}), which is missed in Eq. (\ref{R2}).

In conclusion, let us make  several remarks.
The case of the large $k$-values
(\ref{sol1}) is most interesting because the electron--phonon
 and electrodynamic contributions  into the phonon width
(first and second terms in the parentheses, respectively)
can compete. The result depends
on the Debye screening and the wavevector $k$. In  Raman experiments,
the parameter
$kv_F/|\omega_0+i\gamma|\sim \omega^{(i)}v_F/c\omega_0\simeq 0.3$
if $\omega^{(i)}\simeq 10^4$K and $\omega_0\simeq 10^2$K, and
for metals $v_F\simeq 10^8$ cm/s. Therefore, the high incident-light
frequency or neutron experiments are desirable. It is more simply
to observe the electrodynamic effect in semiconductors, where $k_0$ is
smaller. We have an example of such  experiments in  works \cite{PKC},
where  the metal--insulator
transition was observed in the GaN crystal under pressure.
For the conducting phase,
the longitudinal mode  softens and obtains the additional damping in
comparison with the insulator state. Then, using  Eq. (\ref{sol}), we
calculate \cite{FKC} the collision rate $\gamma$ which is
consistent with the value obtained from the conductivity.

The author thanks E.~G.~Mishchenko for fruitful discussions. 
The work was partially supported by the RFBR
(project 01-02-16211).

\end{document}